# Structural, dielectric and electrocaloric properties of $(Ba_{0.85}Ca_{0.15})(Ti_{0.9}Zr_{0.1-x}Sn_x)O_3$ ceramics elaborated by sol-gel method

S. Belkhadir[a], A. Neqali[a], M. Amjoud[a], D. Mezzane[a], A.Alimoussa[a], E.Choukri[a] ,I.Raevski[b], , Y.Gagou[c,*], M. El Marssi[c], Igor A. Luk'yanchuk[c], Z. Kutnjak[d], B. Rožič[d]

[a] Laboratoire de la matière condensée et nanostructures LMCN, F.S.T.G. Université Cadi Ayyad, BP 549, Marrakech, Morocco

[b] Research Institute of Physics and Faculty of Physics, Southern Federal University, 194 Stachki Ave., 344090, Rostov-on-Don, Russia

[c] Laboratoire de physique de la matière condensée LPMC, Université de Picardie Jules Verne, 33 rue Saint-Leu, 80039, Amiens Cedex, France

[d] Laboratory for calorimetry and dielectric spectroscopy, Condensed Matter Physics Department, Jozef Stefan Institute, Ljubljana, Slovenia.

# Abstract

Ferroelectric ceramics $(Ba_{0.85}Ca_{0.15})(Ti_{0.9}Zr_{0.1-x}Sn_x)O_3$ (x=0.00, 0.02, 0.04, 0.06) were prepared by a sol-gel method. Structural investigation revealed the co-existence of tetragonal (*P4mm*) and orthorhombic (*Pmm*2) symmetries at room temperature for the undoped ceramic, while only a tetragonal structure (*P4mm)* was observed for the doped ceramics. Dielectric measurements indicate a dielectric relaxation process at high temperatures which is essentially related to the hopping of oxygen vacancies $V^{\cdot\cdot}_O$. Furthermore, a down shifting of the Curie temperature ($T_C$) with increasing $Sn^{4+}$ doping rate has been revealed. The temperature profiles of the Raman spectra unveiled the existence of polar nanoregions (PNRs) above the Curie temperature in all ceramics. The ferroelectric properties were found to be related to the microstructure. Electrocaloric effect was investigated in this system that revealed an electrocaloric responsivity of $0.225 \times 10^{-6}$ K m/V for the composition with x = 0.04 Sn doping, where other remarkable physical properties were also observed.

# 1. Introduction:

Lead-zirconate-titanate Pb(Zr$_{1-x}$Ti$_x$)O$_3$ (PZT) ceramics have got diverse functional applications, for instance actuators, sensors, ultrasonic transducers [1–6]. However, despite their distinguished electromechanical performances, these lead-based piezoelectric materials involve more than 60 % of weight with lead, so because of the high toxicity of lead, these materials are potentially detrimental to human health and environment [3,7]. Therefore, the development of new lead-free piezoelectric ceramics becomes imperative.

Barium titanate BaTiO$_3$ (BT) with dopants represents a lead-free alternative and has attracted extensive attention of researchers due to its high electromechanical coupling factor k$_{33}$ ≈0.50, and piezoelectric coefficient (d$_{33}$ ~190 pC/N) [4,8]. Pure BT performances as a piezoelectric material are, however, not matching to those of PZT. Doping A and B-sites with different dopants for instance (A = Ca, Sr, La; B = Nb, Ta, Zr) seems to have a huge effect on enhancement the piezoelectric and dielectric properties of BaTiO$_3$. Recently Liu and Ren studied the effect of Ca and Zr on BT solid solutions and reported an exceptional piezoelectric properties (d$_{33}$~620pC/N) with relatively low Curie temperature (T$_C$~93°C) for xBa(Zr$_{0.2}$Ti$_{0.8}$)O$_3$-(1-x)(Ba$_{0.7}$Ca$_{0.3}$)TiO$_3$ (x =0.5) (BCZT) composition. Besides its remarkable properties as a piezoelectric material, BCZT solid solution system is free from any volatile element and possesses a morphotropic phase boundary (MPB) [9]. BCZT-based ceramics has thus become one of the most suitable lead-free candidates to replace PZT-based materials.

Furthermore, Singh et al. studied the effect of Ca$^{2+}$ on the electrocaloric (EC) properties of (Ba$_{1-x}$Ca$_x$)(Zr$_{0.05}$Ti$_{0.95}$)O$_3$ solid solution, and found a large electrocaloric responsivity 0.253 × 10$^{-6}$ K m /V for the composition (Ba$_{0.92}$Ca$_{0.08}$)(Zr$_{0.05}$Ti$_{0.95}$)O$_3$ [10], which underlined the role of dopants on the electrocaloric properties. For some applications, it is necessary to move T$_C$ to lower temperatures [11,12]. In this respect, it has been observed that the Curie temperature of BaTiO$_3$-based electroceramics is sensitive to tin (Sn$^{4+}$) content [13]. The incorporation of a tin ions in the matrix of Ba(Zr$_{0.1}$Ti$_{0.9}$)O$_3$ demonstrated that the Curie temperature T$_C$ decreases toward room temperature with a large EC responsivity of 0.220 × 10$^{-6}$ K m/V ,this effect could be very desirable for future development of microelectronic cooling applications near the room temperature [14]. Likewise, Patel *et al*. studied the effect of Sn-doping on the electrocaloric performance of BCZT ceramic in a series of Ba$_{0.85}$Ca$_{0.15}$Zr$_{0.1}$Ti$_{0.9-x}$Sn$_x$O$_3$ compositions and reported an enhancement in the electrocaloric properties [15].

Therefore, BCZT ceramics synthesized using wet chemical techniques (such as sol–gel method) as reported by Wang *et al*. showed an excellent electrical performance comparing to the ceramics prepared by solid state method due to the stoichiometric composition of the resultant phase and the chemical purity [16].It is also worth to mention that the electric transport behavior within the ceramic is related to different oxygen content between the grain and the grain boundary, which makes the oxygen vacancies important factor in electric behavior of the ferroelectric material[17,18]. The dielectric relaxation behavior observed at high temperatures for many perovskite titanates such as $(Pb_{0.9}La_{0.1})TiO_3$ [19] ,$Ca_{0.90}Sr_{0.10}Cu_3Ti_{3.95}Zn_{0.05}O_{12}$[20] and $(Sr_{1-1.5x}Bi_x)TiO_3$ [21] is attributed to the hopping of double oxygen vacancies $V_O^{\cdot\cdot}$ [22].

In this work, $(Ba_{0.85}Ca_{0.15})(Ti_{0.9}Zr_{0.1-x}Sn_x)O_3$ (x=0,0.02,0.04,0.06) ceramics were elaborated by using the sol-gel process. A particular emphasis was put on the study of the Sn doping on the microstructure, dielectric, ferroelectric and electrocaloric properties of the $(Ba_{0.85}Ca_{0.15})(Ti_{0.9}Zr_{0.1})O_3$ (BCZT) matrix. The presented results in this paper should provide a better understanding of the relation between the microstructure and the ferroelectric properties. We also expect to achieve a considerable EC effect in a wide temperature range near the room temperature which is very promising for cooling applications.

## 2. Experimental section

$(Ba_{0.85}Ca_{0.15})(Ti_{0.9}Zr_{0.1-x}Sn_x)O_3$ (x=0.00, 0.02, 0.04 and 0.06) ceramics were prepared by employing the sol–gel method. The Barium acetate $(Ba(CH_3COO)_2)$, zirconium oxychloride $(ZrOCl_2 \cdot 8H_2O)$, calcium nitrate tetrahydrate $(Ca(NO_3)_2 \cdot 4H_2O)$, Tin(II) chloride dehydrate $(SnCl_2.2H_2O)$ and titanium isopropoxide $(C_{12}H_{28}O_4Ti)$ were used as initial precursors.

Stoichiometric quantity of $(ZrOCl_2 \cdot 8H_2O)$,$(Ca(NO_3)_2 \cdot 4H_2O)$ and $(SnCl_2.2H_2O)$ were mixed with 2-methoxythanol separately for 1h. Then the required amount of $Ba(CH_3COO)_2$ was dissolved in acetic acid. Finally, the prepared solutions were mixed with $C_{12}H_{28}O_4Ti$.

The obtained precursor solution was heated at 80°C for 1h producing the formation of a viscous yellow color gel. The resulting powder was calcined at 1000 °C for 4h. Then, the pellets pressed at 2.5 ton /cm$^2$ were sintered at 1350 °C during 2h.

The crystalline structure of the ceramics was characterized by X-ray diffraction (XRD), using the Panalytical X-Pert Pro. The measurements were performed at room temperature by using the Cu-Kα radiation with λ ~ 1.540598 Å. The ceramics surface morphology was studied by using a Scanning Electron Microscope (SEM), VEGA 3-Tescan. Raman spectroscopy experiments were carried out on a Horiba Labram HR 800 instrument. The ferroelectric hysteresis loops were obtained by a ferroelectric test system (TF Analyzer 3000, aixACCT). A precision LCR Meter, HP 4284A was used to measure the dielectric properties of the sintered ceramics in the frequency range from 20 Hz to 1 MHz and temperature ranging from 25°C to 450°C.

## 3. Results and discussion

### 3.1 Phase formation and microstructure

Fig. 1 shows the X-ray diffraction (XRD) patterns of the $(Ba_{0.85}Ca_{0.15})(Ti_{0.9}Zr_{0.1})$ powder with different doping levels calcined at 1000 °C for 4h. All samples display a pure perovskite structure ($ABO_3$) aside from any detectable impurity phase within the resolution limit of the used diffractometer, indicating that $Ca^{2+}$, $Zr^{4+}$, and $Sn^{4+}$ have diffused into the $BaTiO_3$ lattice to form a complete solid solution. Furthermore, the calcination temperature of the sol-gel method at 1000°C is significantly lower than 1350°C calcination temperature that would be used in the solid-state synthesis of the same composition [15,23]. The undoped sample $(Ba_{0.85}Ca_{0.15})(Ti_{0.9}Zr_{0.1})O_3$ shows the best refinement fit to a combination of tetragonal (*P*4*mm*) and orthorhombic (*Pmm*2) symmetries coexisting at room temperature. However, the Rietveld analyzed X-ray data for the doped compositions with x=0.02, 0.04 and 0.06 showed a meaningful compromise with the tetragonal structure (*P*4*mm*) only. Structural information such as lattice parameters and atomic positions for all composition are gathered in Table 1. Considering the tolerance factor, one can suppose that small ions ($r(R^{n+}) < 0.87$ Å) will occupy the B-site, large ions ($r(R^{n+}) > 0.94$ Å) will occupy the A-site, and intermediate ions will occupy both sites of the perovskite structure with different ratios [24]. When taking into account $Ba^{2+}$(r = 1.35 Å), $Ca^{2+}$(r = 1 Å), $Zr^{4+}$(r = 0.72 Å), $Ti^{4+}$(r = 0.605 Å), $Sn^{4+}$(r = 0.69 Å), it is clear that $Sn^{4+}$ ions should occupy B site as $(Zr, Ti)^{4+}$ that was confirmed by structural refinements. Fig. 2 shows the variation of the unit cell volume calculated for the tetragonal structure for all samples. We observed a decrease in the volume values with increasing of $Sn^{4+}$ doping as a consequence of the substitution of $Zr^{4+}$ (r = 0.72 Å) by a smaller ion radius $Sn^{4+}$(r = 0.69 Å).

Fig. 3 shows scanning electron micrographs of $(Ba_{0.85}Ca_{0.15})(Ti_{0.9}Zr_{0.1-x}Sn_x)O_3$ ceramics sintered at 1350°C. The effect of $Sn^{4+}$ is highly noticeable: the doped ceramics seem to have a larger grain size that increases gradually from 1.5 μm for the undoped ceramic

$(Ba_{0.85}Ca_{0.15})(Ti_{0.9}Zr_{0.1})O_3$ to 5μm for the doped ceramic with x=0.04. The higher grain growth rate observed could be related to the active diffusion of $Sn^{4+}$ due to its smaller ionic radii (r=0.69Å) as compared to $Zr^{4+}$ (r=0.72 Å) during sintering. However, with increasing $Sn^{4+}$ doping concentration toward x=0.06, the porosity increases as a consequence of the accumulation of $Sn^{4+}$ ions at the grain boundaries when the concentration level exceeds the solubility limits [4]. The density values of the doped ceramics with x=0.00, 0.02, 0.04 and 0.06 are found to be 5.45 , 5.72, 5.32 and 4.80 g/cm$^3$ respectively.

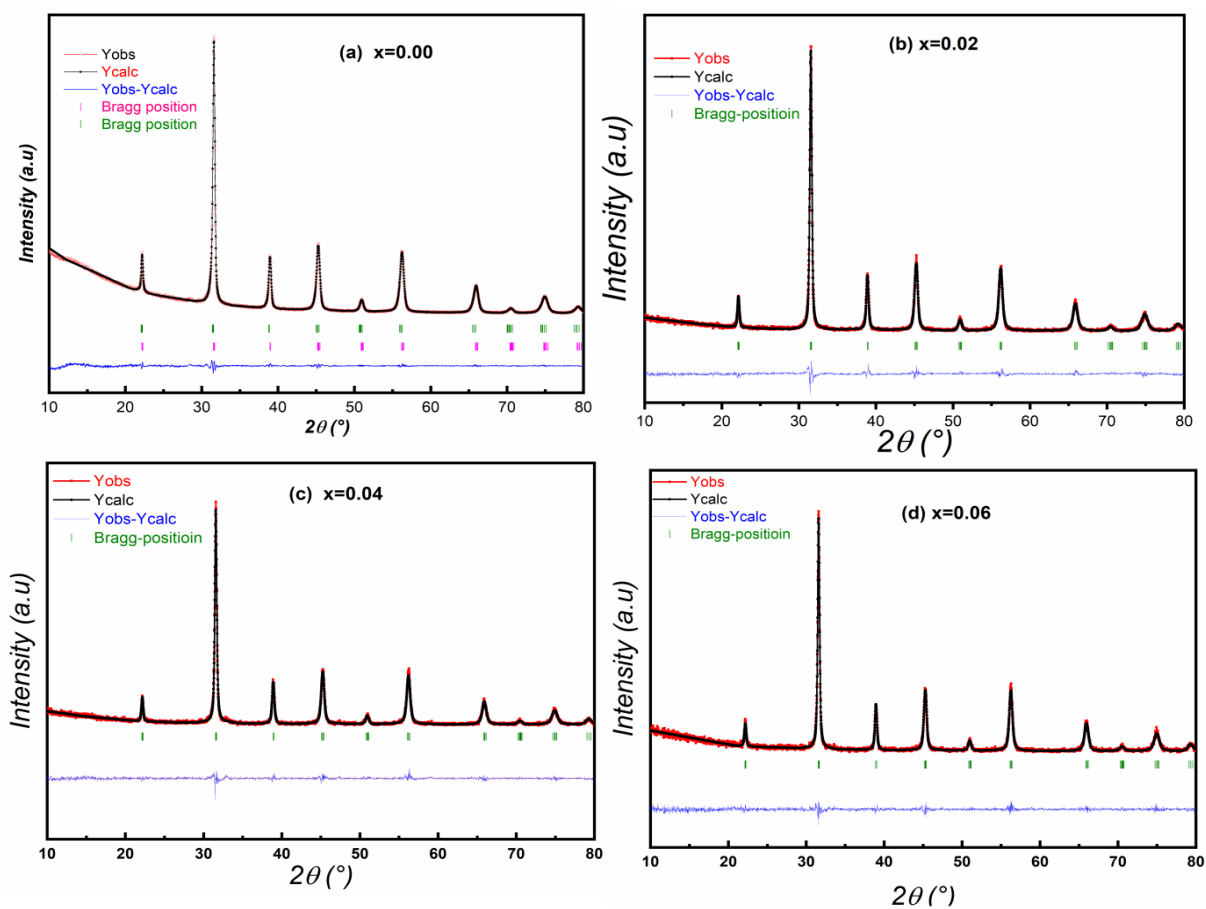

Fig.1. Rietveld fitted X-ray diffraction patterns of $(Ba_{0.85}Ca_{0.15})(Ti_{0.9}Zr_{0.1-x}Sn_x)O_3$ powders for (a) x=0.00, (b) x=0.02, (c) x=0.04, (d) x=0.06

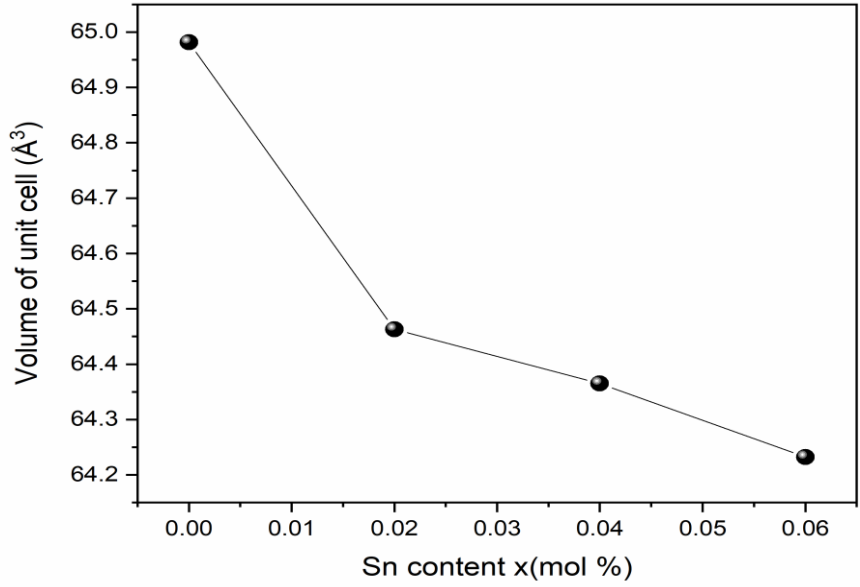

Fig.2. Variation of the unit cell volume as a function of Sn content.

Table 1: Structural parameters of $(Ba_{0.85}Ca_{0.15})(Ti_{0.9}Zr_{0.1-x}Sn_x)O_3$ obtained from Rietveld refinement.

| | | | | | | |
|---|---|---|---|---|---|---|
| | x=0.00 | | | | | |
| Phase | Tetragonal | | | Orthorhombic | | |
| Space group | P4mm | | | Pmm2 | | |
| | x | y | z | x | y | z |
| Ba/Ca | 0.00000 | 0.00000 | 0.06036 | 0.00000 | 0.00000 | -0.00668 |
| Ti/Zr | 0.50000 | 0.50000 | 0.59931 | 0.50000 | 0.50000 | 0.45863 |
| $O_1$ | 0.50000 | 0.00000 | 0.38910 | 0.50000 | 0.00000 | 0.44133 |
| $O_2$ | 0.50000 | 0.50000 | 0.12395 | 0.50000 | 0.50000 | 0.08662 |
| $O_3$ | - | - | - | 0.00000 | 0.50000 | 0.54699 |
| Lattice parameters(Å) | a = b = 4.0285  c = 4.0041 (% phase = 40.62 ) | | | a = 3.9945, b = 4.0105 c = 4.0070 (% phase=59.38) | | |
| | x=0.02 | | | | | |
| | Tetragonal P4mm | | | | | |
| | x | | y | | z | |
| Ba/Ca | 0.00000 | | 0.00000 | | -0.13204 | |
| Ti/Zr/Sn | 0.50000 | | 0.50000 | | 0.38906 | |
| $O_1$ | 0.50000 | | 0.50000 | | 0.83082 | |
| $O_2$ | 0.00000 | | 0.50000 | | 0.34624 | |
| Lattice parameters(Å) | a =b=4.0038 | | | | c = 4.0213 | |
| | x=0.04 | | | | | |
| | Tetragonal P4mm | | | | | |
| | x | | y | | z | |
| Ba/Ca | 0.00000 | | 0.00000 | | -0.02591 | |
| Ti/Zr/Sn | 0.50000 | | 0.50000 | | 0.49072 | |
| $O_1$ | 0.50000 | | 0.50000 | | 0.01774 | |
| $O_2$ | 0.00000 | | 0.50000 | | 0.46649 | |
| Lattice parameters(Å) | a =b=4.0026 | | | | c = 4.0176 | |
| | x=0.06 | | | | | |
| | Tetragonal P4mm | | | | | |
| | x | | y | | z | |
| Ba/Ca | 0.00000 | | 0.00000 | | -0.02752 | |

| | | | |
|---|---|---|---|
| Ti/Zr/Sn | 0.50000 | 0.50000 | 0.49362 |
| $O_1$ | 0.50000 | 0.50000 | -0.03284 |
| $O_2$ | 0.00000 | 0.50000 | 0.49006 |
| Lattice parameters(Å) | a =b=4.0002 | | c = 4.0141 |

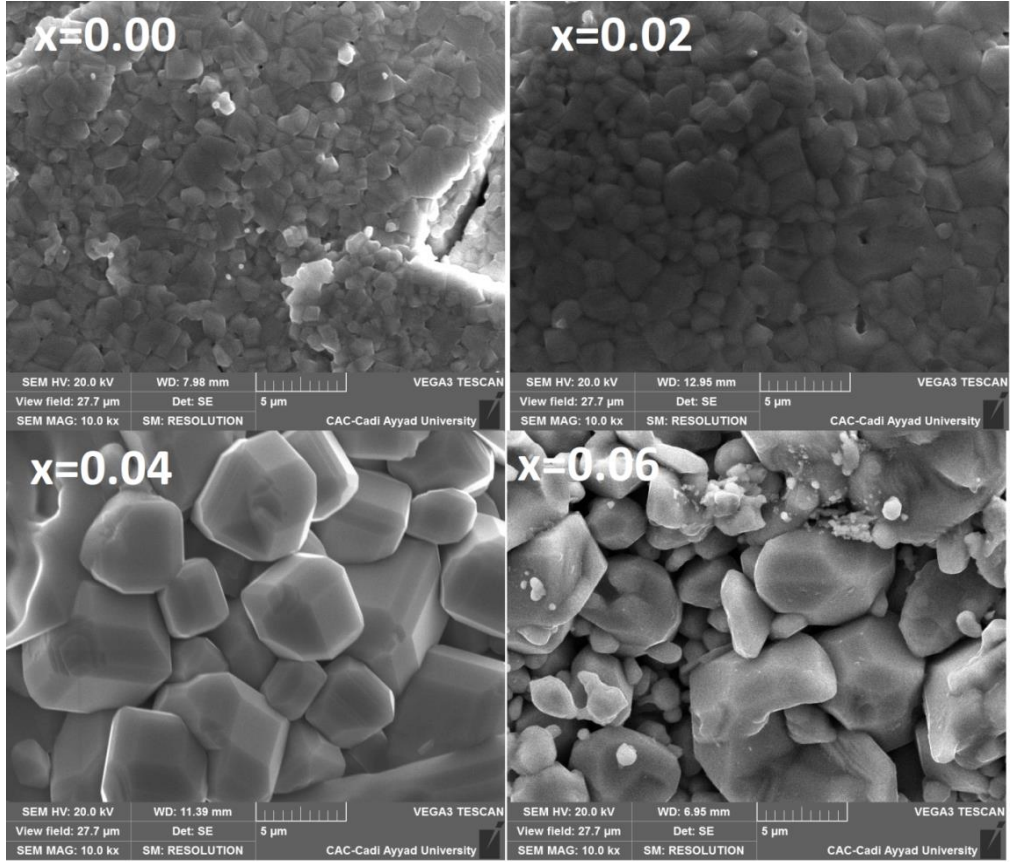

Fig.3. SEM micrographs of $(Ba_{0.85}Ca_{0.15})(Ti_{0.9}Zr_{0.1-x}Sn_x)O_3$ ceramics sintered at 1350°C for 2h.

## 3.2 Dielectric studies

Fig. 4(a-d) display the temperature dependence of the dielectric constant ($\varepsilon_r$) measured at various frequencies for the undoped sample $(Ba_{0.85}Ca_{0.15})(Ti_{0.9}Zr_{0.1})O_3$ (BCZT). The dielectric anomaly related to the orthorhombic-to-tetragonal (O–T) phase transition at room temperature (~298K) corroborates the coexistence of orthorhombic and tetragonal phases in the ceramic at room temperature. The dielectric anomaly related to the tetragonal-to-cubic (T–C) phase transition at 369 K signals the ferroelectric-paraelectric (FE-PE) structural change for the undoped sample. When $Sn^{4+}$ doping rate increases, the Curie temperature shifts down to 343 K, as shown in Fig. 5. This effect can probably be due to the formation of non-ferroelectric $SnO_6$ octahedra that interrupt the bonding between the ferroelectrically active $TiO_6$ octahedra

[25]. With the addition of $Sn^{4+}$, we notice the existence of only one phase transition (T–C), as a result of O–T transition shifting toward lower temperatures. Owing to its low $T_C$, this family ceramics might be appropriate for near-room-temperature applications.

We noticed also that the value of the dielectric constant increases with increasing Sn content even with the reduced density. Figure 6 illustrates the dielectric losses (tanδ) for all compositions as a function of temperature measured at 1 kHz. At the room temperature the dielectric losses decrease except for compositions x=0.04 and 0.06, which could be related to the low density of these samples.

To analyze the dielectric relaxation observed at around 225°C-425°C, the temperature and the frequency (*f*) response of dielectric loss have been plotted for samples doped x=0.00 and 0.02 (Fig 7 (a),(b), (c), (d),(e) and (f) ). The existence of loss peaks in the dielectric spectrum with low frequency dispersion at high temperatures reveals the contribution of localized relaxation and non-localized charge transport processes. Figs. 7 (a-f) show the variation of loss factor for the compositions x=0.0 and x=0.02, as a function of temperature and *f*. Loss factor is higher for low frequencies and its maximum decreases and deviate toward elevated temperature when frequency increases, as shown in Figs. 7(a) and 7(b). Moreover, loss factor exhibits higher value for high temperature in the FE and PE phases while the opposite occurs around Curie temperature as seen in Figs. 7(c) and 7(d). All this observations are in favor of ionic conduction via hopping mechanism. An identical dielectric relaxation behavior has been described in other papers for many perovskites [21,26–33].

The activation energy of the dielectric relaxation was calculated using the Debye-like relation described as:

$$f = f_0 \exp\left(-\frac{E_{relax}}{k_B T_m}\right) , \qquad (1)$$

Where $E_{relax}$ is the activation energy for relaxation, $k_B$ is the Boltzmann's constant, *f* is the frequency of the applied ac field, i.e., the inverse of the experimental time scale, $f_0$ is the pre-exponential factor and $T_m$, the temperature at which the dielectric loss reaches a maximum.

The obtained $E_{relax}$ after fitting the experimental data (Fig 6.(e) and 6(f)) using Eq. (1) is 1.02 eV for the pure $(Ba_{0.85}Ca_{0.15})(Ti_{0.9}Zr_{0.1})O_3$ and 0.98 eV for the sample with x=0.02. These values correspond to the activation energy of the motion of the doubly ionized oxygen vacancies ($V_O^{\cdot\cdot}$) [17,34]. Consequently, the observed dielectric relaxation process for the samples is governed by the hopping of $V_O^{\cdot\cdot}$.

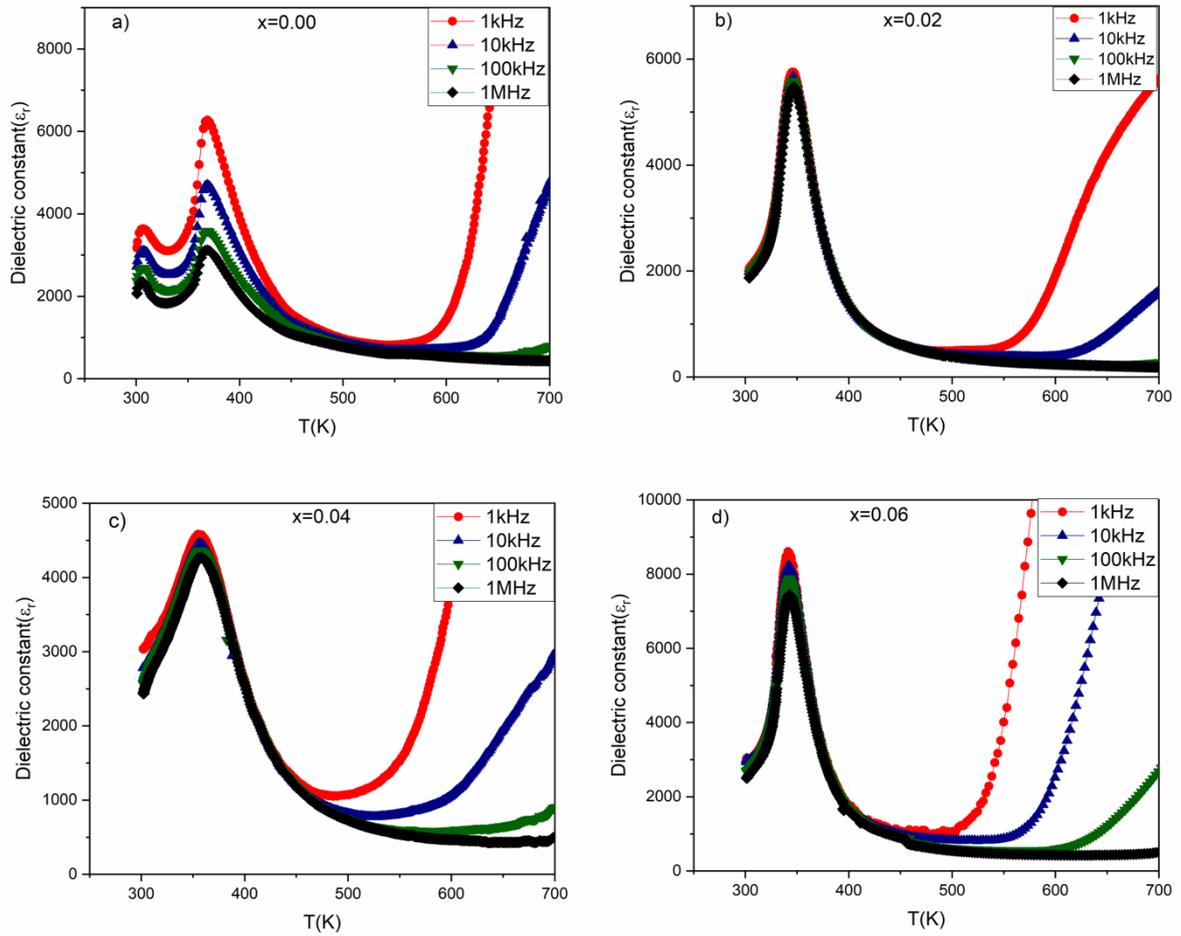

Fig. 4. Temperature dependence of the dielectric constant for $(Ba_{0.85}Ca_{0.15})(Ti_{0.9}Zr_{0.1-x}Sn_x)O_3$, (a) x=0.00, (b) x=0.02, (c) x=0.04, and (d) x=0.06.

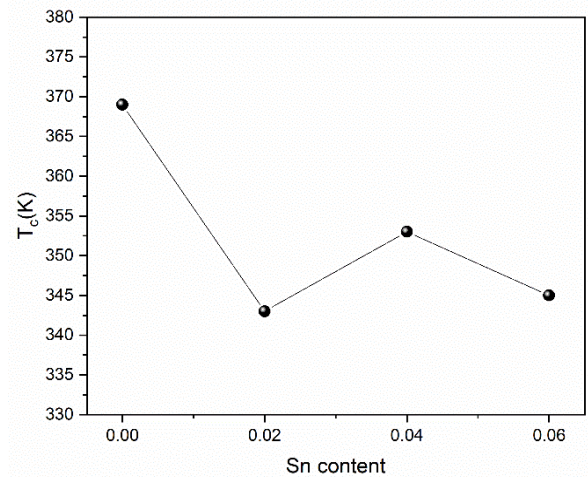

Fig.5. Variation of the Curie temperature as a function of Sn content.

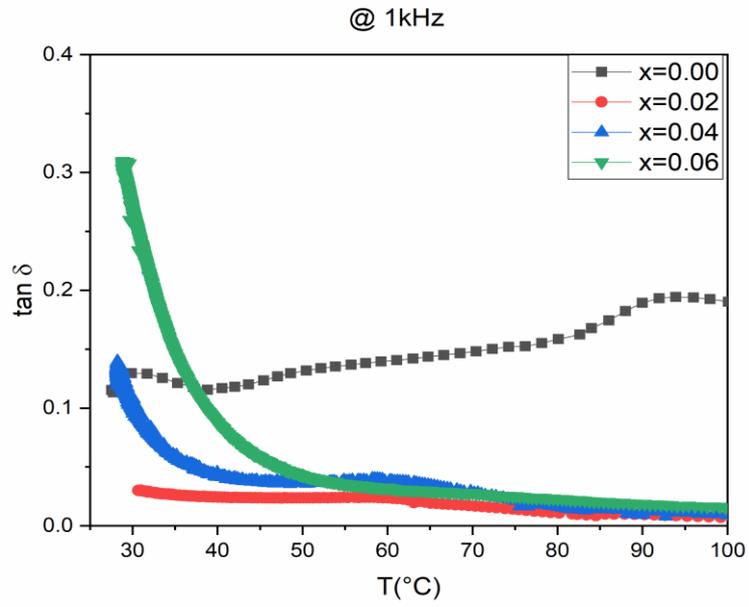

Fig.6.The evolution of the dielectric losses as a function of temperature for all composition at 1kHz.

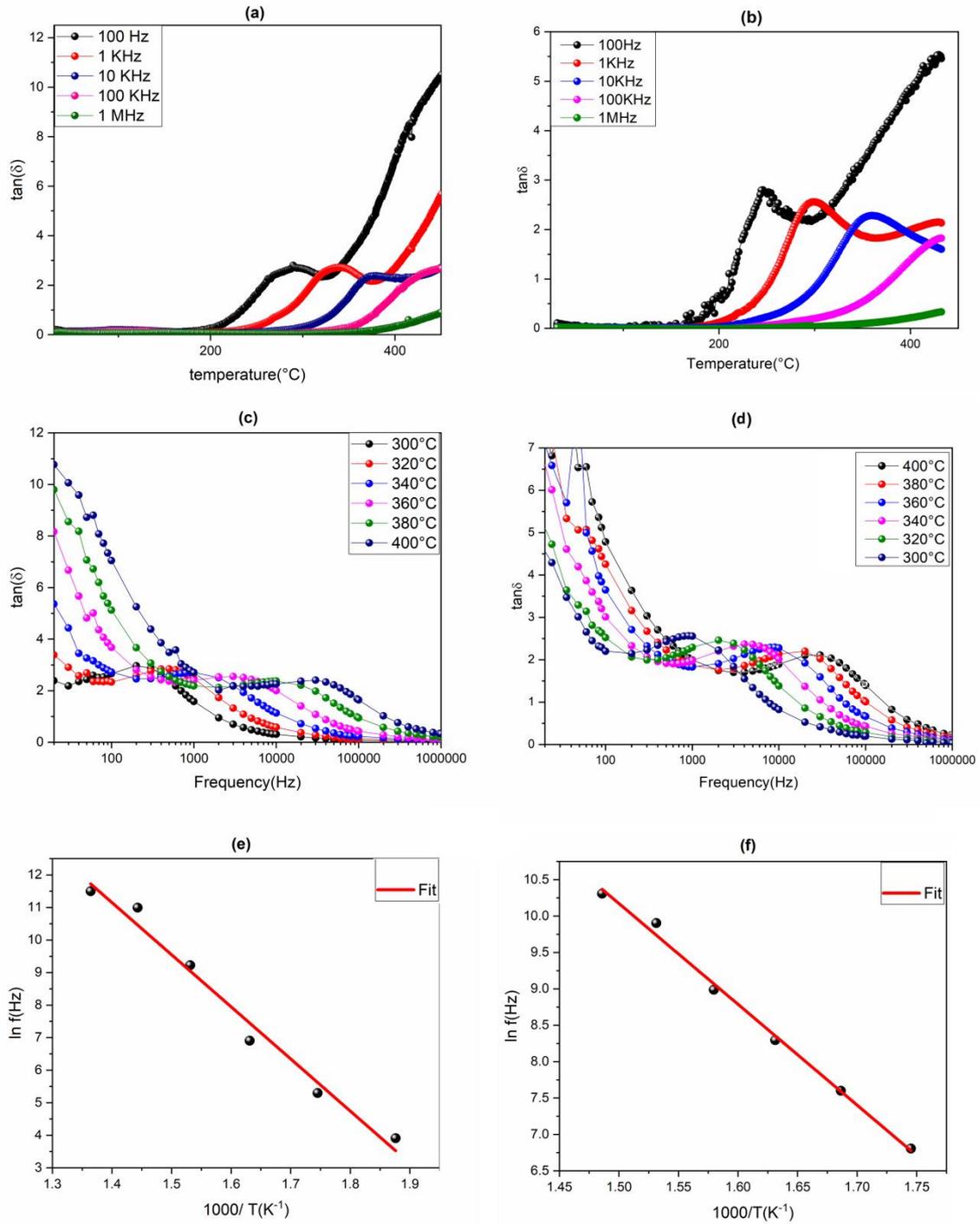

Fig.7. Temperature dependence of the loss tangent measured at various frequencies for ceramics (a)x=0.00,(b) x=0.02, Frequency dependence of tan δ at various temperatures for ceramics (c)x=0.00,(d) x=0.02, variation of the frequency corresponding to the tanδ(f) maximum as a function of the inverse of temperature.(e) x=0.00 ,(f) x=0.02.

## 3.3 Raman spectroscopic studies

The ferroelectric phase transition and the local structure can be investigated by the Raman spectroscopy considering its sensitivity to crystal lattice vibration and ferroelectric phase transition [35–37]. Fig.8. shows Raman spectra recorded at room temperature for different compositions. For the all doped ceramics, two transverse optical modes [$A_1$ ($TO_1$)] and [$A_1$ ($TO_2$)] were detected in the range of 100 - 250 $cm^{-1}$. Another band at around 300 $cm^{-1}$ assigned to a combined mode [$B_1$, E (TO + LO)] is also considered as a particular distinctive of the tetragonal phase. The Raman mode assignments are considered to be the same as those for pure $BaTiO_3$ [38,39], whereas a clear mode was labeled at 206 $cm^{-1}$ for the pure $(Ba_{0.85}Ca_{0.15})(Ti_{0.9}Zr_{0.1})O_3$ (BCZT) and reported to be assigned to the $A_1(LO)$ mode [40,41]. This band is reported to be more noticeable in the Rhombohedral (R) and Orthorhombic(O) phases, and it disappears in the tetragonal phase [40–42], which suggests the co-existence of orthorhombic and tetragonal phases in accordance with the X-ray diffraction data. The thermal evolution of the Raman spectra has been presented for compositions x = 0.00 and 0.02 in Fig. 9. The ferroelectric (tetragonal)-to-paraelectric (cubic) transition for all compositions can be recognized by the weakening of both modes at 300 $cm^{-1}$ and 720 $cm^{-1}$. The variation of the full width at half maxima (FWHM) for the composition x=0.02 is shown in Fig. 10. We notice a change in the slope at about 70°C, associated with the phase transition which further supports the phase transition indicated by the dielectric measurements. However, the two strongest Raman modes, at 522 $cm^{-1}$(A1/E(TO)) and 720 $cm^{-1}$(A1/E(LO)) remain in the cubic phase, this can be related to the existence of ferroelectric clusters or polar nano-regions (PNRs) [35,43–45].

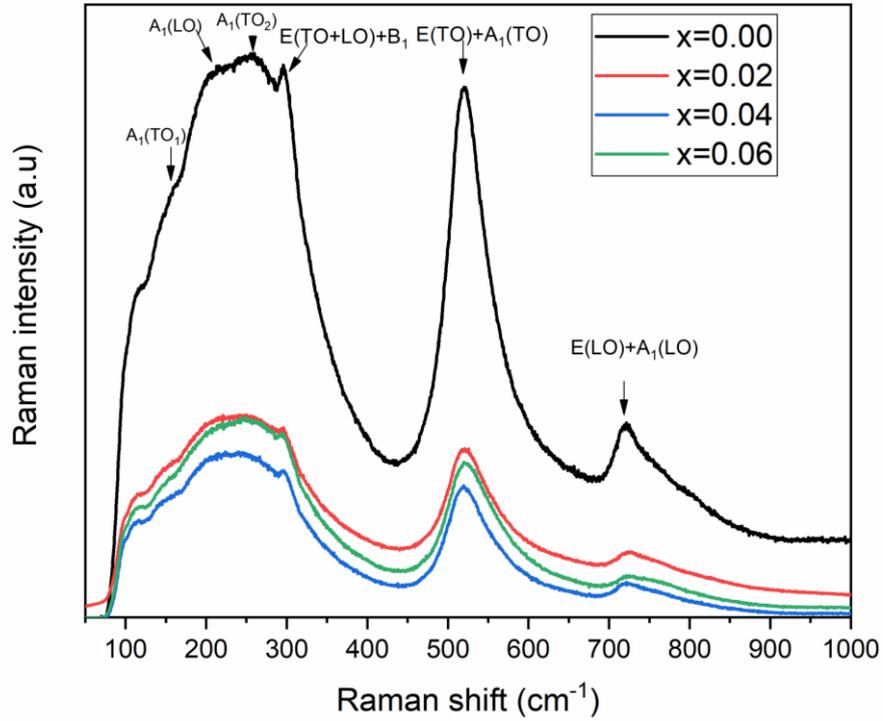

Fig.8. Room-temperature Raman spectra of $(Ba_{0.85}Ca_{0.15})(Ti_{0.9}Zr_{0.1-x}Sn_x)O_3$ ceramics.

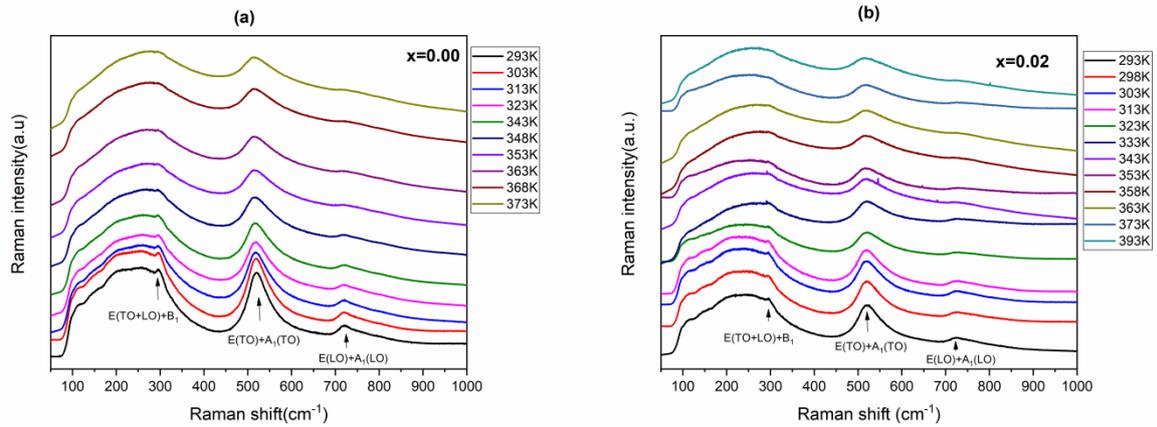

Fig. 9. Temperature dependence of Raman spectra for $(Ba_{0.85}Ca_{0.15})(Ti_{0.9}Zr_{0.1-x}Sn_x)O_3$ x=0.00 (a) and x=0.02 (b) ceramics.

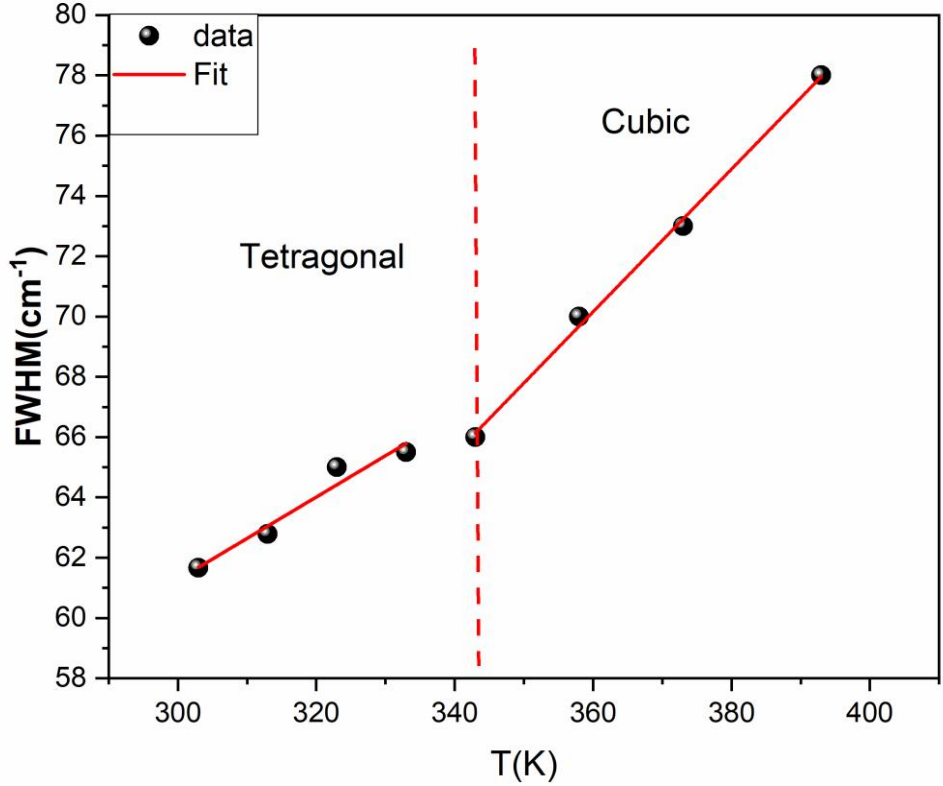

Fig. 10. Temperature dependence of the full width at half maximum (FWHM) for $A_1/E(TO)$ band centered at 520 cm$^{-1}$ for doped ceramic with x=0.02.

## 3.4 Ferroelectric studies

P-E hysteresis loops for $(Ba_{0.85}Ca_{0.15})(Ti_{0.9}Zr_{0.1-x}Sn_x)O_3$ (x=0.00, 0.02, 0.04, 0.06) measured at room temperature are shown in Fig.11(a). The saturated P-E loops indicate the ferroelectric character of the samples. It can be seen that the remnant polarization ($P_r$) and the coercive field ($E_C$) are highly influenced by Sn doping rate. Both exhibit a similar behavior, the remnant polarization increases up to a value of $2P_r \sim 9.7$ μC/cm$^2$ observed at x=0.04, pursued then by a reduction with increasing the doping values as shown in Fig 11(b) and (c). This variation is significantly related to the grain size of the ceramics. Orihara has proposed the equation (2) based on a modified Kolmogorov-Avrami model [46]

$$f_{ds} = f_{ds\,0}\left[1 - \exp(\frac{-G_a d^3}{kT})\right] \qquad (2)$$

Where $f_{ds}$ is the domain switchability, $f_{ds\,0}$ is the initial polarization or domain of ferroelectric materials, $G_a$ is a constant that represents grain anisotropy energy density, d is a grain size, k is a constant rate, and T is temperature. It is clear from the equation (2) that the increase in the grain size is followed by an easier domain wall rotation due to the raise of the domain switchability which ameliorate the ferroelectric properties [47]. Similar results were reported by Buatip *et al*. [48] . The low ferroelectric properties observed for the sample with x=0.06 could be due to the increase of porosity which decreases dramatically the density of the sample.

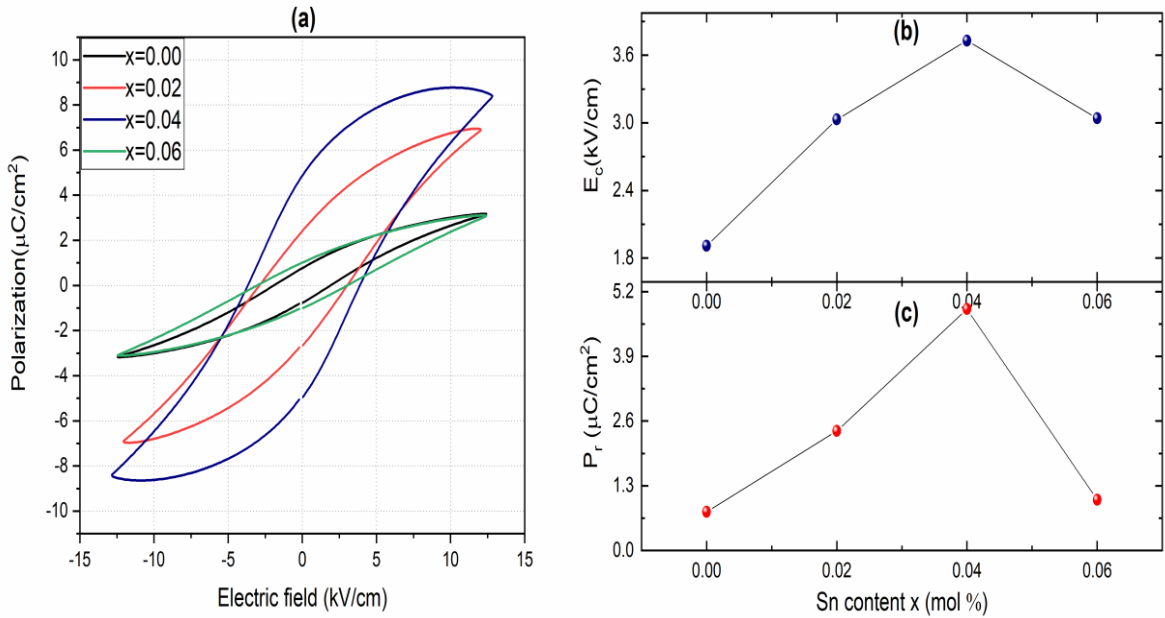

Fig.11. (a) Room temperature P–E curves $(Ba_{0.85}Ca_{0.15})(Ti_{0.9}Zr_{0.1-x}Sn_x)O_3$ ceramics, (b) variation of coercive field $E_C$ and (c) remnant polarization $P_r$ with Sn content

## 3.5 Electrocaloric properties

In order to evaluate the electrocaloric effect (ECE), P–E hysteresis loops were performed as a function of temperature for all samples. Fig. 12 shows the thermal variation for P-E for the composition x=0.02. Polarization gradually decreases when temperature increases due to the thermal agitation which progressively destroys the dipoles order arrangement and led to phase transition.

The ECE adiabatic temperature change (ΔT) was calculated by indirect method using Maxwell's relation $(\partial P/\partial T)_E = (\partial S/\partial E)_T$, that permits to obtain the following equation:

$$\Delta T = -\frac{1}{\rho}\int_{E_1}^{E_2} \frac{T}{C_p}\left(\frac{\partial P}{\partial T}\right)_E dE , \qquad (3)$$

Where $\rho$ is the density of the sample, and $C_p$ is the specific heat assumed to have a constant value for all ceramics Cp ~ 0.35 J/gK [49].

Fig. 13 shows the adiabatic variation of the electrocaloric temperature change (ΔT) for several electric fields and for all the compositions. The observed anomalies are induced by the phase transitions which are considerably enhanced by the applied electric field. It is also worth to mention that the broad peak of the temperature change (ΔT) observed for all samples gives a larger temperature window which is advantageous for refrigeration applications.

Fig. 14 shows the temperature change (ΔT) as a function of Sn-content for an applied electric field 12 kV/cm. It can be seen that the (ΔT) is increasing with increasing the Sn content, in a similar manner as the ferroelectric properties, achieving a value of 0.27 K under 12 kV/cm electric field for the composition x=0.04 at 342 K, near the room temperature. The calculated electrocaloric temperature change (ΔT) from the indirect method corresponds to EC responsivity $\zeta = 0.225\times10^{-6}$ K m/V. This value is close to that reported by Patel et al. for the composition $Ba_{0.85}Ca_{0.15}Zr_{0.1}Ti_{0.88}Sn_{0.02}O_3$ [15]. Hence, it may be inferred that Sn doping can significantly enhance the EC properties and lower the Curie temperature which is in a positive agreement with the dielectric study.

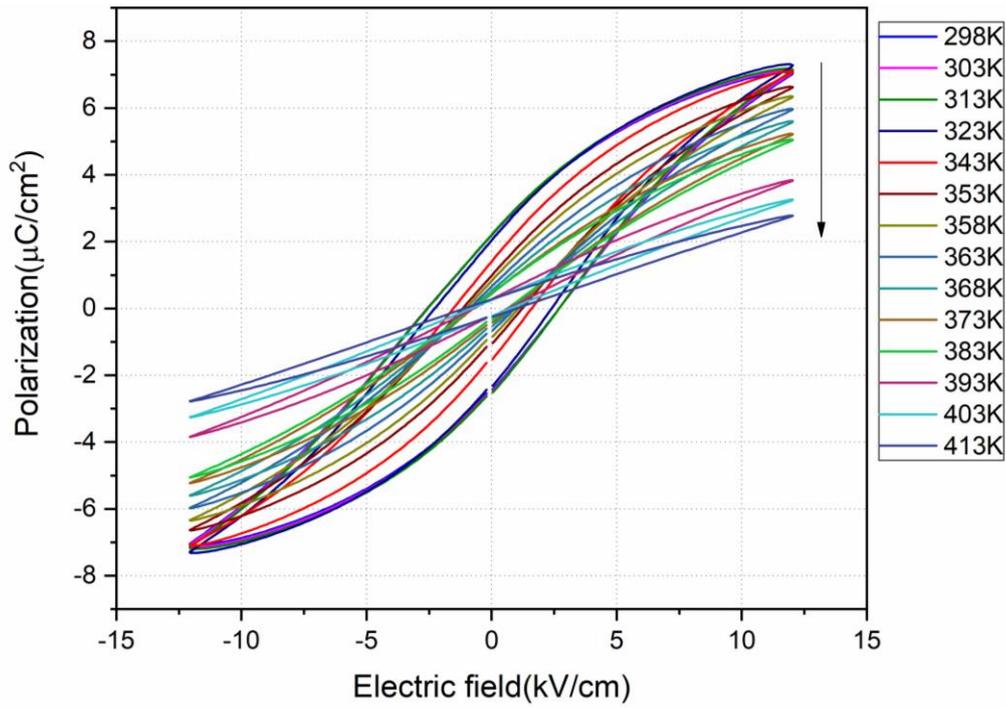

Fig 12. Thermal variation of P-E hysteresis loops for doped ceramic with x=0.02.

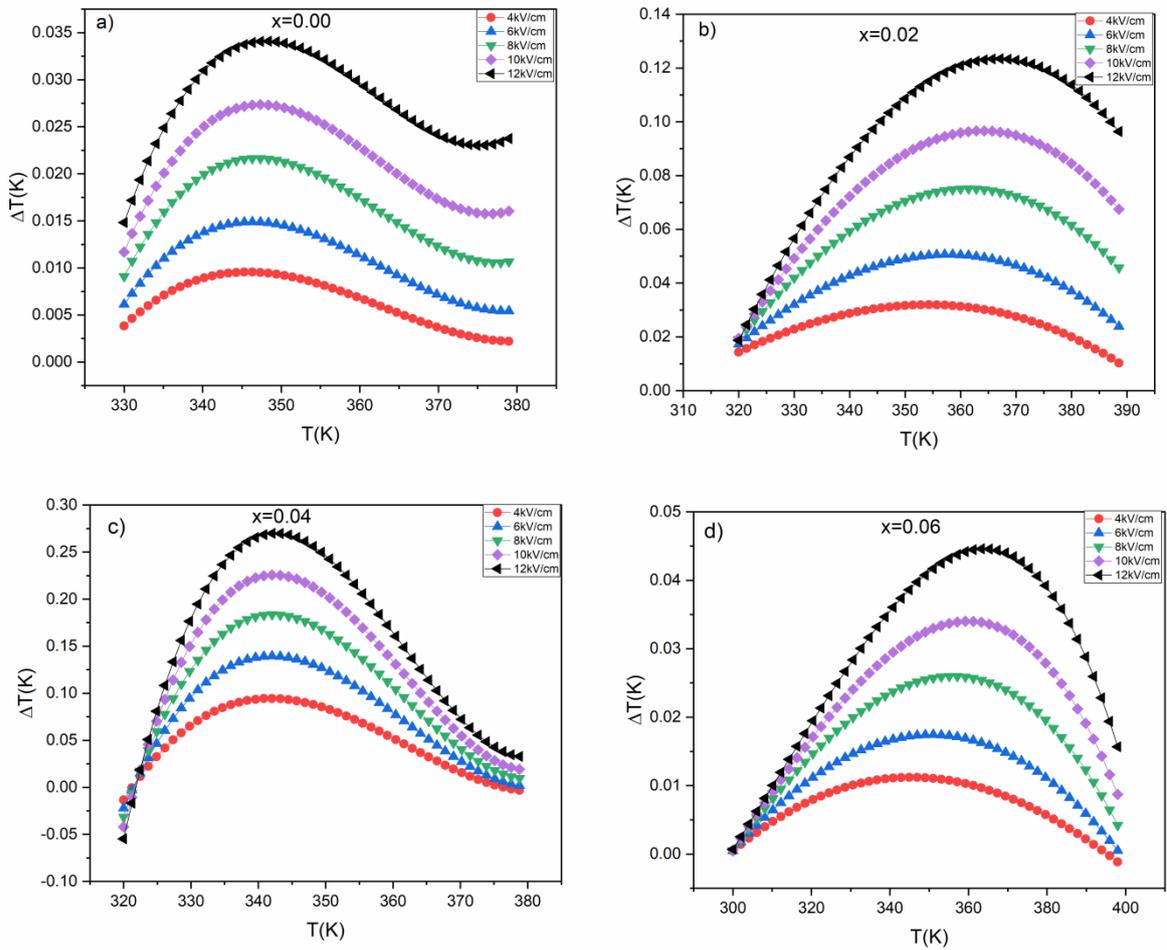

Fig. 13.Electrocaloric temperature change (ΔT) as a function of temperature at different applied field for $(Ba_{0.85}Ca_{0.15})(Ti_{0.9}Zr_{0.1-x}Sn_x)O_3$, (a) x=0.00, (b) x=0.02, (c) x=0.04, and (d) x=0.06.

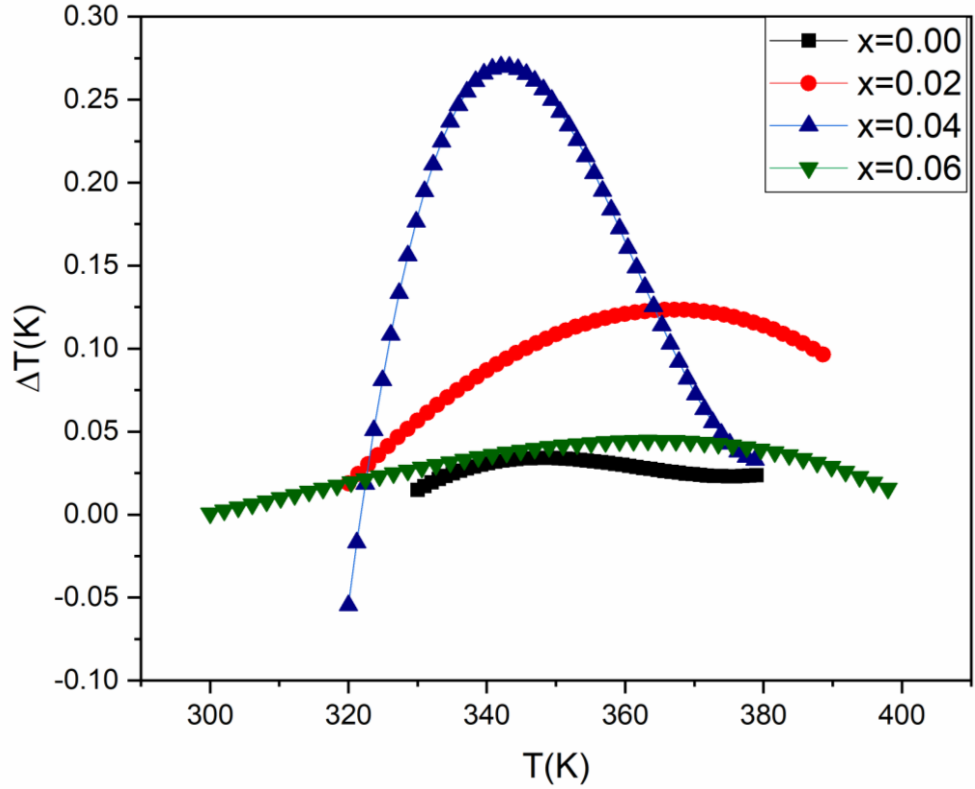

Fig. 14. Temperature profiles of the electrocaloric temperature change (ΔT) for various Sn content at 12 kV/cm applied electric field.

# Conclusions

Effect of Sn doping on lead free $(Ba_{0.85}Ca_{0.15})(Ti_{0.9}Zr_{0.1})O_3$ ceramics has been investigated. The samples were successfully synthesized by sol-gel method at low temperature of 1000 °C. The X-ray diffraction patterns and Raman spectra analysis show the co-existence of tetragonal and orthorhombic symmetry for the undoped sample at room temperature. A noticeable decrease of the Curie temperature ($T_C$) has been observed as a result of doping the $(Ba_{0.85}Ca_{0.15})(Ti_{0.9}Zr_{0.1})O_3$ host matrix with $Sn^{4+}$. The dielectric relaxation process observed at high temperatures has been found to be essentially due to the hopping of $V_O^{\cdot\cdot}$. Ferroelectric properties are significantly enhanced by Sn-substitution and a maximum remnant polarization value ($2P_r \sim 9.7 \mu C/cm^2$) was observed for the ceramic with x=0.04. Electrocaloric study highlights an influential improvement in EC properties by Sn-doping. The highest value of ΔT = 0.27 K under 12 kV/cm applied electric field was achieved for x=0.04 which corresponds to an EC responsivity of $\zeta = 0.225 \times 10^{-6}$ Km/V. This EC responsivity rivals those observed in lead-based materials such as PMN-PT, PZT and PLZT materials, thus making Sn doped lead-free $(Ba_{0.85}Ca_{0.15})(Ti_{0.9}Zr_{0.1})O_3$ ceramics a good candidate for electrocaloric cooling applications.

# Acknowledgements

The authors gratefully acknowledge the financial support of CNRST Priority Program PPR 15/2015 and the European Union's Horizon 2020 research and innovation program ENGIMA under the Marie Skłodowska-Curie grant agreement No 778072. I.R. acknowledges the support from the Ministry of Education and Science of the Russian Federation via the project 3.1649.2017/4.6. B.R. and Z.K. acknowledge support from Slovenian Research Agency project J1-9147 and program P1-0125.